\documentclass[conference]{IEEEtran}
\IEEEoverridecommandlockouts
\usepackage{array}
\usepackage[caption=false,font=normalsize,labelfont=sf,textfont=sf]{subfig}
\usepackage{textcomp}
\usepackage{stfloats}
\usepackage{url}
\usepackage{verbatim}
\usepackage{graphicx}
\usepackage{cite}
\usepackage[utf8]{inputenc}
\usepackage{booktabs}
\usepackage{amsmath,amssymb,amsfonts}
\usepackage{amsthm}
\usepackage{algorithm}
\usepackage{algpseudocode}
\usepackage{tablefootnote}
\usepackage{graphicx}
\usepackage{epsfig}
\usepackage{tikz}
\usepackage{float}
\usepackage{xcolor}
\usepackage{multirow}
\usepackage{adjustbox}
\usepackage{stfloats}
\usepackage{url}
\usepackage{balance}
\usepackage[verbose]{geometry}  
\begin{document}

\title{RIS-Assisted Survivable Backhaul Recovery in Small-Cell Systems}

\author{

 \IEEEauthorblockN{Zhenyu Li$^*$, \"Ozlem Tu\u{g}fe Demir$^\dagger$, Emil Bj{\"o}rnson$^*$, Cicek Cavdar$^*$}
\IEEEauthorblockA{ {$^*$Department of Computer Science, KTH Royal Institute of Technology, Stockholm, Sweden}
\\ 
 {$^\dagger$Department of Electrical and Electronics Engineering, Bilkent University, Ankara, Turkiye}
\\
		{Email: zhenyuli@kth.se, ozlemtugfedemir@bilkent.edu.tr, emilbjo@kth.se, cavdar@kth.se }
}

\thanks{This study is conducted under the Eureka Celtic Project RAI-6Green: Robust and AI Native 6G Green Mobile Networks (C2023/1-9) and partly supported by Swedish Wireless Innovations Center: SweWIN (2023-00572), both funded by Swedish Innovation Agency Vinnova.}
}



\maketitle

\begin{abstract}
    The increasing densification of small-cell networks substantially expands cable-based backhaul infrastructure, creating heightened vulnerability to cable link failures. This paper proposes a reconfigurable intelligent surface (RIS)-assisted backup framework that exploits a key insight: during backhaul cable failures, base station (BS) radio components remain functional, enabling wireless backhaul traffic redistribution. Our framework maintains network connectivity by redistributing disconnected BS backhaul traffic to neighboring BSs through RIS-assisted wireless links. To maximize survivability across varying traffic conditions, we formulate a joint optimization problem that maximizes total resolvable backhaul traffic by jointly deciding BS selection, RIS phase shifts, and precoding vectors. The inherent non-convexity arising from coupling and quadratic fractional term is addressed through an alternating optimization algorithm that iteratively solves tractable convex subproblems via quadratic transformation. Comprehensive numerical evaluations demonstrate that the proposed RIS-enhanced framework significantly improves survivability from 58\% to 72\% under challenging high-intensity hotspot traffic conditions. Moreover, RIS provides the greatest gains for antenna-constrained systems by extending coverage to access more spare capacity of the distant BSs as well as enhancing the signal strength. Consequently, high survivability is achieved even with only two antennas per BS under moderate traffic intensity.
\end{abstract}


\section{Introduction}

    The exponential mobile data growth drives network densification with small-cell deployments relying on cable-based backhaul. However, cable failures create critical vulnerability, occurring at rates of 0.7 to 2 per 100 miles annually~\cite{epri2023underground}. Caused by construction, natural disasters, aging, and attacks~\cite{gallagher2023undersea}, these failures disconnect hundreds of users and contribute to 20 billion dollars in annual operator outage costs~\cite{donegan2013mobile}. Enhancing network survivability against backhaul failures is thus essential.
    
    Recognizing these vulnerabilities, existing approaches to enhance network survivability can be categorized into proactive and reactive strategies. Proactive approaches build resilience before failures through redundant connectivity paths~\cite{11046130} or network function virtualization~\cite{6979954}, but require substantial infrastructure investment that is economically infeasible for dense small-cell deployments, and virtualization cannot overcome physical link failures. Reactive approaches respond after failures by deploying emergency infrastructure~\cite{10685402, 9599638}, but are generally designed for catastrophic scenarios requiring extensive deployment time, making them unsuitable for rapid response to routine localized disruptions. A critical gap remains for cost-effective solutions that can rapidly restore connectivity by leveraging existing wireless infrastructure, without requiring extensive upfront investment or emergency deployment procedures.

    Given that radio components typically remain functional during cable failures, wirelessly redistributing backhaul traffic emerges as a promising backup solution. While wireless backhaul traditionally faces capacity and environmental challenges~\cite{townend2023challenges}, dense small-cell deployments fundamentally alter this landscape through shorter propagation distances, thus larger inter-BS wireless capacity. Moreover, reconfigurable intelligent surfaces (RISs) are expected to be integrated into dense networks to enhance transmission quality and overcome the signal blockage during normal operation~\cite{shi2024ris,huang2021towards,zhang2021reconfigurable}, meaning RIS infrastructure is already economically justified with no additional deployment costs for survivability. Crucially, RISs maintain operationality during cable failures, unlike conventional network components that depend on backhaul connectivity, enabling dynamic adaptation for emergency traffic redistribution. While extensive research demonstrates RIS capabilities for transmission enhancement, their potential for network survivability remains largely unexplored. Survivability refers to the capability to maintain service continuity and quality when certain failures happen~\cite{10187713}. Motivated by these points, this work investigates RIS-assisted survivable backhaul recovery for small-cell networks. The main contributions of this work are summarized as follows:
    \begin{itemize}
        \item We propose a novel RIS-assisted backhaul traffic redistribution framework that maintains connectivity under a single backhaul connection failure by redistributing disconnected BS backhaul traffic to neighboring BSs via RIS-assisted wireless links.
        
        \item We develop a joint optimization algorithm that alternately optimizes redistributing target selection, RIS phase-shift configurations, and precoding vectors, effectively handling the non-convexity from coupled binary and continuous variables.
        
        \item Numerical results demonstrate that RIS assistance significantly enhances survivability under a high-intensity hotspot traffic, with the most pronounced gains observed in antenna-constrained scenarios.
    \end{itemize}

\section{System model}

    \begin{figure}[tb]
        \centering
        \includegraphics[trim= 210 100 220 110, clip, width=0.9\linewidth]{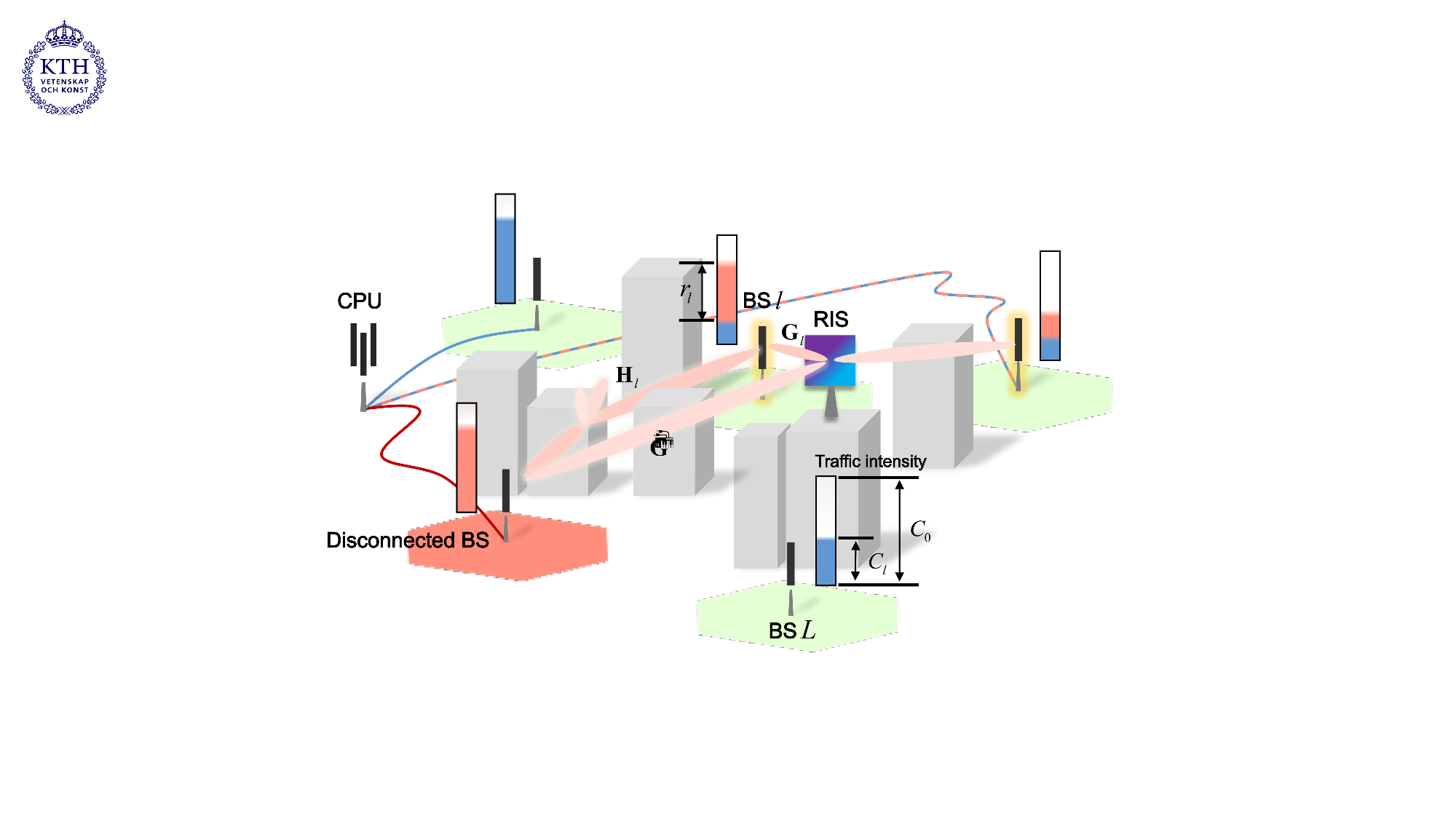}
                \vspace{-2mm}
        \caption{Illustration of the RIS-assisted survivable backhaul recovery in a small-cell system.}
        \vspace{-4mm}
        \label{fig:systemmodel}
    \end{figure}
    
    As illustrated in Fig.~\ref{fig:systemmodel}, we consider a small-cell cluster where multiple BSs, each equipped with $N$ antennas, transmit backhaul traffic to a central processing unit (CPU) via dedicated cable connections. When a cable fails, a varying number of BSs will be disconnected depending on the network topology and failure location. Given the high reliability of cable connections, we focus on single-link failures. To minimize disruption impact, we adopt a star topology where each BS connects directly to the CPU, ensuring that any single failure disconnects exactly one BS, which we refer to as the \emph{disconnected BS}. The remaining $L$ \emph{surviving BSs} are indexed by $l\in\{1,2,\ldots,L\}$.

    The BSs are having baseband units (BBUs) with a designed capability of processing a traffic rate of $C_0$. Before the link failure, each BS $l$ has a different local traffic load $C_l$ due to the different user behavior in its serving cell. After the link failure, the local traffic $C_d$ of the disconnected BS needs to be resolved to survive the failure. Since the BBU processing capacity is provisioned for peak traffic loads that are rarely experienced in practice~\cite{barlacchi2015multi, 8466626},  a spare capacity $C_0-C_l>0$ typically exists under normal operating conditions. The disconnected BS can leverage this spare capacity by wirelessly redistributing its local traffic to neighboring BSs via decode-and-forward, enabling indirect connection to the core network. Given the temporal and spatial variation of the amount of spare capacity, selection of the traffic-distributing BSs, and the corresponding channel condition, the achievable total redistributable traffic $R$ can be different. Referring to~\cite{10187713}, we define the \emph{survivability} against single backhaul link failure as
    \begin{equation}
        \psi = \min\left(100\%,R/C_d~|~C_0,\left\{C_l\right\}_{l=1}^L,C_d\right) ,\label{eq:survivability}
    \end{equation}
    where $R$ is the resolvable backhaul traffic. To maximize the survivability, an RIS that consists of $M$ reflecting elements, closest to the disconnected BS, is considered to enhance the reachability and the transmitted data stream quality. Following a failure event, an emergency protocol is activated to enable BSs to exchange local traffic conditions and to facilitate remote reconfiguration of the RIS. For simplicity, perfect channel state information is assumed to be available at all BSs. The BSs and RIS will jointly perform backhaul traffic redistribution based on this.  

    \subsection{RIS-assisted emergent traffic redistribution channel}
        
        To effectively redistribute the backhaul traffic, multi-user transmission is employed. The spatial multiplexing constraint limits the number of parallel streams to at most $N$, thereby constraining the number of traffic-redistributing BSs to $N$. Since optimal redistribution depends on both available spare capacity and channel quality, the suitable receiving BSs are not necessarily the closest ones geographically. We introduce a binary redistributing target selection indicator $\beta_l\in\{0, 1\}$, where $\beta_l=1$ indicates that surviving BS $l$ is selected to receive redistributed traffic.
    
        To highlight the survivability enhancement effect, the emergent traffic redistribution is carried on a reserved orthogonal band to the access link, for instance, the mmWave band. Moreover, an independent power budget $P_0$ is applied to perform this operation. Under these assumptions, the effective disconnected BS-RIS-surviving BS $l$ channel, denoted as $\mathbf{H}_{\text{eff}, l}\in\mathbb{C}^{N\times N}$, is given as
        \begin{equation}
            \mathbf{H}_{\text{eff},l}(\mathbf{\Phi})=\mathbf{H}_l+\mathbf{G}_l\mathbf{\Phi}\tilde{\mathbf{G}},
        \end{equation}
        where $\mathbf{H}_l\in\mathbb{C}^{N\times N}$, $\tilde{\mathbf{G}}\in\mathbb{C}^{M\times N}$, and $\mathbf{G}_l \in\mathbb{C}^{N\times M}$ represent the channel matrices between the disconnected BS and surviving BS $l$, the disconnected BS and RIS, and the RIS and surviving BS $l$, respectively. Moreover, $\mathbf{\Phi}\in\mathbb{C}^{M\times M}$ is a diagonal matrix that contains the phase shifts applied by the RIS. Furthermore, $\phi_m$ is the $m$-th diagonal element of $\mathbf{\Phi}$.

        We denote the precoder applied for the transmission between the disconnected BS and the surviving BS $l$ as $\mathbf{w}_l\in\mathbb{C}^{N\times 1}$. If minimum-mean-square-error (MMSE) combining is utilized at the surviving BS $l$, the maximum traffic that can be resolved by BS $l$ can be expressed as 
        \begin{equation}
            r_l = B\log_2(1+\mathbf{w}_l^H\mathbf{H}_{\text{eff},l}(\mathbf{\Phi})^H\mathbf{R}_l^{-1}\mathbf{H}_{\text{eff},l}(\mathbf{\Phi})\mathbf{w}_l) ,\label{eq:individualrate}
        \end{equation}
        where $B$ is the available bandwidth and $\mathbf{R}_l \in \mathbb{C}^{N\times N}$ is the interference plus noise covariance term for the surviving BS $l$ when treating all unintended signals as noise. It is expressed as 
        \begin{equation}
            \mathbf{R}_l = \sigma^2\mathbf{I}_N+\sum_{j\neq l}\mathbf{H}_{\text{eff},l}(\mathbf{\Phi})\mathbf{w}_j\mathbf{w}_j^H\mathbf{H}_{\text{eff},l}(\mathbf{\Phi})^H,
        \end{equation}
        where $\sigma^2$ is the additive noise variance. The total redistributable traffic becomes $R=\sum_{l=1}^L\beta_lr_l$.
    
\section{RIS-enhanced traffic redistributing}

    The following optimization problem is formulated to maximize the survivability of the disconnected BS by configuring the backhaul traffic distributing target selection, precoder, and RIS phase shifts: 
    \begin{subequations}
        \begin{align}
            \textbf{P0:}\quad\underset{\{\mathbf{w}_l, \beta_l, \phi_m\}}{\text{maximize}} \quad &\psi \label{eq:achievablerload}\\
            \text{subject to} \quad 
            &  \beta_lr_l\leq (C_0 - C_l), \quad \forall l ,\label{eq:sparecapacity} \\
            & \sum_{l=1}^L \beta_l \leq N ,\label{eq:spatialstream}\\
            & \|\mathbf{w}_l\|^2 \leq \beta_l P_{\max},\quad \forall l, \label{eq:powerselection}\\
            & \sum_{l=1}^{L}\|\mathbf{w}_l\|^2 \leq P_{\max}, \label{eq:precoderpower}\\
            & |\phi_m| =1,\quad \forall m,\label{eq:modulusphase}\\
            & \beta_l \in \{0, 1\},\quad \forall l.\label{eq:binary}
        \end{align}
    \end{subequations}
    The objective function~\eqref{eq:achievablerload} seeks to maximize the survivability $\psi$. Constraint~\eqref{eq:sparecapacity} ensures that the redistributed traffic does not exceed the available spare capacity of each selected BS. Constraint~\eqref{eq:spatialstream} restricts the maximum number of spatial data streams to not exceed the number of transmit antennas at the disconnected BS. Constraint~\eqref{eq:powerselection} enforces zero power allocation for non-selected BSs. Constraint~\eqref{eq:precoderpower} maintains the total transmit power of the disconnected BS within the power budget. The lossless reflection property of the RIS is captured by constraint~\eqref{eq:modulusphase}, which restricts each reflecting element to have unit modulus. Finally, constraint~\eqref{eq:binary} enforces the binary nature of the BS selection variables.

    Problem \textbf{P0} is non-convex due to the coupling between $\mathbf{w}_l$ and $\boldsymbol{\Phi}$ in the rate expressions, and the bilinear coupling between binary variables $\beta_l$ and rates $r_l$. In the remainder of this section, we propose an alternating optimization framework to iteratively optimize $\mathbf{w}_l$ and $\boldsymbol{\Phi}$.

    \subsection{RIS-enhanced traffic redistributing algorithm}

        To assess the system's maximum capability in redistributing backhaul traffic, we employ an epigraph reformulation of $\textbf{P0}$ where the objective becomes maximizing $R$ subject to the constraint
        \begin{equation}
            \sum_{l=1}^L\beta_lr_l \geq R. \label{eq:originalepi}
        \end{equation}
        Given the survivability definition in~\eqref{eq:survivability}, maximizing $R$ is equivalent to maximizing $\psi$ under any system condition. This reformulation leads to an equivalent problem while improving mathematical tractability. 
        
        Subsequently, to address the bilinear coupling between $\beta_l$ and $r_l$ in \textbf{P0}, auxiliary variables $f_l\in\mathbb{R}_+$ are introduced to replace the bilinear terms through the following constraint set:
        \begin{align}
            &f_l\leq r_l+(1-\beta_l)\mathcal{M},\label{eq:bilinear1}\\
            &f_l \leq \beta_l\mathcal{M},\label{eq:bilinear2}
        \end{align}
        where $\mathcal{M}$ represents a sufficiently large positive constant. Through this substitution, constraints~\eqref{eq:originalepi} and~\eqref{eq:sparecapacity} are equivalently transformed into
        \begin{align}
            &\sum_{l=1}^L f_l \geq R, \label{eq:auxsumload}\\
            &f_l \leq C_0-C_l,\quad \forall l. \label{eq:auxsparecapacity}
        \end{align}
        
        Having resolved the bilinear coupling issue, the coupling between the precoding vectors $\mathbf{w}_l$ and the RIS phase-shift matrix $\mathbf{\Phi}$ within the rate expression $r_l$ is addressed next. For mathematical convenience, \eqref{eq:individualrate} is reformulated using the equivalent epigraph constraint as
        \begin{equation}
            B\log_2(1+\mathbf{w}_l^H\mathbf{H}_{\text{eff},l}(\mathbf{\Phi})^H\mathbf{R}_l^{-1}\mathbf{H}_{\text{eff},l}(\mathbf{\Phi})\mathbf{w}_l) \geq r_l,\quad \forall l. \label{eq:individualrateepi}
        \end{equation}
        Employing the quadratic transform approach~\cite{shen2018fractional}, constraint~\eqref{eq:individualrateepi} is further reformulated as
        \begin{align}
            &-\left(\mathbf{y}_l^{(\varepsilon)}\right)^H\left(\sigma^2\mathbf{I}_N+\sum_{j\neq l}\mathbf{H}_{\text{eff},l}(\mathbf{\Phi})\mathbf{w}_j\mathbf{w}_j^H\mathbf{H}_{\text{eff},l}(\mathbf{\Phi})^H\right)\mathbf{y}_l^{(\varepsilon)} \nonumber\\
            &\hspace{4mm} +2\Re\left\{\left(\mathbf{y}_l^{(\varepsilon)}\right)^H\mathbf{H}_{\text{eff},l}(\mathbf{\Phi})\mathbf{w}_l\right\}\geq t_l,\quad \forall l,\label{eq:quadratictransform}\\
            & \hspace{4mm} t_l \geq 2^{r_l/B}-1,\quad \forall l, \label{eq:exponentialcone}
        \end{align}
        where $\mathbf{y}_l^{(\varepsilon)}\in\mathbb{C}^{N\times 1}$ denotes the auxiliary variable computed from the outputs of the $(\varepsilon-1)$-th iteration, and $t_l\in\mathbb{R}_+$ is an auxiliary variable introduced to formulate the exponential cone constraint in~\eqref{eq:exponentialcone}. Note that the coupling between $\mathbf{w}_l$ and $\mathbf{\Phi}$ persists in constraint~\eqref{eq:quadratictransform}. To decouple these variables, an alternating optimization framework is adopted wherein each variable set is optimized while treating the other as fixed. Following this approach, constraint~\eqref{eq:quadratictransform} becomes convex with respect to either optimization target.
        
        For notational clarity, the auxiliary constant terms corresponding to the cases where $\mathbf{\Phi}$ and $\mathbf{w}_l$ are held fixed are denoted as $\mathbf{y}_{\mathbf{\Phi},l}$ and $\mathbf{y}_{\mathbf{w},l}$, respectively. By substituting $\mathbf{y}_l$ with $\mathbf{y}_{\mathbf{\Phi},l}$ in constraint~\eqref{eq:quadratictransform}, the tractable optimization subproblem for a given $\mathbf{\Phi}$ is formulated as
        \begin{subequations}
            \begin{align}
                \textbf{P1: }&\underset{R,\{\mathbf{w}_l, \beta_l, r_l, t_l, f_l\}}{\text{maximize}} \quad R \\
                &\text{subject to}  \quad 2\Re\left\{\left(\mathbf{y}_{\mathbf{\Phi},l}^{(\varepsilon)}\right)^H\mathbf{H}_{\text{eff},l}(\mathbf{\Phi}^{(\varepsilon)})\mathbf{w}_l\right\}\nonumber\\
                &\hspace{4mm}-\left(\mathbf{y}_{\mathbf{\Phi},l}^{(\varepsilon)}\right)^H\left(\sigma^2\mathbf{I}_N+\sum_{j\neq l}\mathbf{H}_{\text{eff},l}(\mathbf{\Phi}^{(\varepsilon)})\mathbf{w}_j\right.\nonumber\\
                &\hspace{4mm} \left.\cdot\mathbf{w}_j^H\mathbf{H}_{\text{eff},l}(\mathbf{\Phi}^{(\varepsilon)})^H\right)\mathbf{y}_{\mathbf{\Phi},l}^{(\varepsilon)}\geq t_l,\quad \forall l,\\
                &\eqref{eq:spatialstream},~\eqref{eq:powerselection},~\eqref{eq:precoderpower},~\eqref{eq:binary},~\eqref{eq:bilinear1},~\eqref{eq:bilinear2},~\eqref{eq:auxsumload},~\eqref{eq:auxsparecapacity},~\eqref{eq:exponentialcone},\nonumber
            \end{align}
        \end{subequations}
        where $\mathbf{H}_{\text{eff},l}(\mathbf{\Phi}^{(\varepsilon)})$ represents the effective cascaded channel from the disconnected BS through the RIS to the surviving BS $l$, evaluated using the phase configuration $\mathbf{\Phi}^{(\varepsilon)}$. Upon solving subproblem $\textbf{P1}$ at iteration $\varepsilon$, the auxiliary term $\mathbf{y}_{\mathbf{w},l}$ is updated according to
        \begin{align}
            &\mathbf{y}_{\mathbf{w},l}^{(\varepsilon+1)}=\left(\sigma^2\mathbf{I}_N+\sum_{j\neq l}\mathbf{H}_{\text{eff},l}(\mathbf{\Phi}^{(\varepsilon)})\mathbf{w}_j^{(\varepsilon+1)}\left(\mathbf{w}_j^{(\varepsilon+1)}\right)^H\right.\nonumber\\
            &\hspace{4mm}\left.\cdot\mathbf{H}_{\text{eff},l}(\mathbf{\Phi}^{(\varepsilon)})^H\right)^{-1}\mathbf{H}_{\text{eff},l}(\mathbf{\Phi}^{(\varepsilon)})\mathbf{w}_l^{(\varepsilon+1)}+\epsilon\mathbf{1}_N\label{eq:WMMSEupdate1},
        \end{align}
        where $\epsilon$ is a small regularization parameter that ensures numerical stability and maintains exploration of non-selected BSs, improving convergence behavior. Through iterative updates of the auxiliary term via \eqref{eq:WMMSEupdate1}, the corresponding rate $r_l$ is maximized. Given constraint~\eqref{eq:originalepi}, this approach is equivalent to maximizing the objective function, thereby ensuring consistency in the optimization goal.
        
        When the precoding vectors $\mathbf{w}_l$ are held fixed, the non-affine equality constraint~\eqref{eq:modulusphase} brings non-convexity to optimizing $\boldsymbol{\Phi}$. To address this, the constraint is relaxed to an inequality constraint
        \begin{equation}
            |\phi_m| \leq 1,\quad \forall m. \label{eq:relaxedphase}
        \end{equation}
        In the context of maximizing $R$, the magnitudes $|\phi_m|$ will naturally be pushed to the constraint boundary, thereby satisfying the original equality constraint. Similarly, by replacing $\mathbf{y}_l$ with $\mathbf{y}_{\mathbf{w},l}$ in constraint~\eqref{eq:quadratictransform}, the tractable optimization subproblem for given precoding vectors $\mathbf{w}_l$ is formulated as
        \begin{subequations}
            \begin{align}
                \textbf{P2: }&\underset{R,\{\phi_m, \beta_l, r_l, t_l, f_l\}}{\text{maximize}} \quad R \\
                &\text{subject to}  \quad 2\Re\left\{\left(\mathbf{y}_{\mathbf{w},l}^{(\varepsilon+1)}\right)^H\mathbf{H}_{\text{eff},l}(\mathbf{\Phi})\mathbf{w}_l^{(\varepsilon+1)}\right\}\nonumber\\
                &\hspace{4mm}-\left(\mathbf{y}_{\mathbf{w},l}^{(\varepsilon+1)}\right)^H\left(\sigma^2\mathbf{I}_N+\sum_{j\neq l}\mathbf{H}_{\text{eff},l}(\mathbf{\Phi})\mathbf{w}_j^{(\varepsilon+1)}\right.\nonumber\\
                &\hspace{4mm} \left.\cdot\left(\mathbf{w}_j^{(\varepsilon+1)}\right)^H\mathbf{H}_{\text{eff},l}(\mathbf{\Phi})^H\right)\mathbf{y}_{\mathbf{w},l}^{(\varepsilon+1)}\geq t_l,\quad \forall l,\\
                &\eqref{eq:spatialstream},~\eqref{eq:binary},~\eqref{eq:bilinear1},~\eqref{eq:bilinear2},~\eqref{eq:auxsumload},~\eqref{eq:auxsparecapacity},~\eqref{eq:exponentialcone},~\eqref{eq:relaxedphase}.\nonumber
            \end{align}
        \end{subequations}
        
        Following the solution of subproblem $\textbf{P2}$ at iteration $\varepsilon$, the auxiliary variable $\mathbf{y}_{\mathbf{\Phi},l}$ is updated as
        \begin{align}
            &\mathbf{y}_{\mathbf{\Phi},l}^{(\varepsilon+1)}=\left(\sigma^2\mathbf{I}_N+\sum_{j\neq l}\mathbf{H}_{\text{eff},l}(\mathbf{\Phi}^{(\varepsilon+1)})\mathbf{w}_j^{(\varepsilon+1)}\left(\mathbf{w}_j^{(\varepsilon+1)}\right)^H\right.\nonumber\\
            &\hspace{4mm}\left.\cdot\mathbf{H}_{\text{eff},l}(\mathbf{\Phi}^{(\varepsilon+1)})^H\right)^{-1}\mathbf{H}_{\text{eff},l}(\mathbf{\Phi}^{(\varepsilon+1)})\mathbf{w}_l^{(\varepsilon+1)}+\epsilon\mathbf{1}_N.\label{eq:WMMSEupdate2}
        \end{align}
        
        The complete alternating RIS-enhanced load redistribution algorithm is summarized in Algorithm~\ref{alg:risenhancedalgorithm}.
        
        \begin{algorithm}[!t]
            \caption{RIS-Assisted Traffic Redistribution Algorithm} \label{alg:risenhancedalgorithm} 
            \begin{algorithmic}[1]
                \State \textbf{Input:} Channel matrices $\mathbf{H}_{l}$, $\mathbf{G}_{l}$, and $\tilde{\mathbf{G}}$; maximum available capacity $C_0$ and local traffic $C_l$ for all BSs
                \State \textbf{Initialization:} Initialize $\mathbf{\Phi}^{(0)}$ randomly while maintaining $|\phi_m^{(0)}|=1$ for all $m$; initialize $\mathbf{y}_{\mathbf{\Phi},l}^{(0)}$  to ensure feasibility of $\textbf{P1}$; set iteration counter $\varepsilon=0$ and maximum iterations $E$; set regularization parameter $\epsilon = 10^{-6}/\sqrt{N}$
                \While {$\varepsilon < E$}
                    \State Solve $\textbf{P1}$ and update $\mathbf{w}_l^{(\varepsilon)}$ to $\mathbf{w}_l^{(\varepsilon+1)}$
                    \State Update $\mathbf{y}_{\mathbf{w},l}^{(\varepsilon)}$ to $\mathbf{y}_{\mathbf{w},l}^{(\varepsilon+1)}$ according to~\eqref{eq:WMMSEupdate1}
                    \State Solve $\textbf{P2}$ and update $\mathbf{\Phi}^{(\varepsilon)}$ to $\mathbf{\Phi}^{(\varepsilon+1)}$
                    \State $\phi^{(\varepsilon+1)}_m \gets \phi^{(\varepsilon+1)}_m/|\phi^{(\epsilon+1)}_m|\quad\forall m$
                    \State Update $\mathbf{y}_{\mathbf{\Phi},l}^{(\varepsilon)}$ to $\mathbf{y}_{\mathbf{\Phi},l}^{(\varepsilon+1)}$ according to~\eqref{eq:WMMSEupdate2}
                    \State $\varepsilon \gets \varepsilon+1$
                \EndWhile
                \State\textbf{Output:} Optimal BS selection indicators $\beta_{l}^{(E)}$, RIS phase-shift configuration $\mathbf{\Phi}^{(E)}$, and precoding vectors $\mathbf{w}_l^{(E)}$ for the disconnected BS
            \end{algorithmic}
        \end{algorithm}
        Both $\textbf{P1}$ and $\textbf{P2}$ are in the form of mixed-integer programming (MIP) problems, which are NP-hard. However, except for the integer constraints, each sub-problem has a convex form. The integer constraints are addressed using branch-and-bound methods with the commercial solver MOSEK~\cite{mosek}. The branch-and-bound procedure contributes to the dominant computational complexity, which grows exponentially with the number of binary variables in the worst case.

    \section{Simulation setup and results}

        We adopt a realistic channel model reflecting small-cell propagation characteristics. Direct BS-to-BS channels are modeled as non-line-of-sight (NLOS) due to obstruction in dense urban environments, while the RIS is positioned strategically to maintain line-of-sight (LOS) with all BSs.

        The direct channel is $\mathbf{H}_l = \sqrt{\beta^\text{NLOS}(f_c,d_l)}\mathbf{H}_l^\text{small}$ with uncorrelated Rayleigh small-scale fading $[\mathbf{H}_l^\text{small}]_{i,j}\sim\mathcal{CN}(0,1)$, while the RIS-related channels $\mathbf{G}_l=\sqrt{\beta^\text{LOS}(f_c,D_l)}\mathbf{G}_l^\text{small}$ and $\tilde{\mathbf{G}}=\sqrt{\beta^\text{LOS}(f_c,\tilde{d})}\tilde{\mathbf{G}}^\text{small}$ exhibit Rician fading with factor $\kappa$. Here, $d_l$, $D_l$, and $\tilde{d}$ denote the distances between disconnected BS and surviving BS $l$, RIS and surviving BS $l$, disconnected BS and RIS, respectively, and $f_c$ is the carrier frequency. Large-scale fading follows the UMi model with isotropic antennas~\cite{sun2016propagation}, and BSs are deployed on a hexagonal grid with inter-site distance $d_{0}$\,m. The RIS closest to the disconnected BS is reconfigured for backhaul recovery, and for simplicity, we assume the RIS is transparent to the signal~\cite{mu2021simultaneously}, neglecting its angular attenuation. Parameters are summarized in Table~\ref{tab:coefficient}. 

        \begin{figure}[tb]
            \centering
            \includegraphics[width=0.75\linewidth]{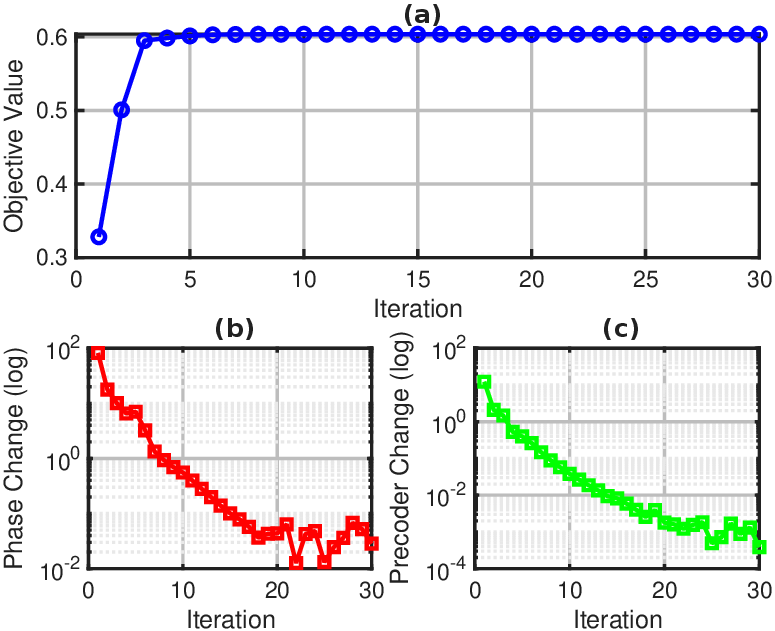}
            \caption{Convergence condition of the algorithm. (a) Objective value changing over iterations; (b) phase shifts changing over iterations; (c) precoders changing over iterations.}
            \label{fig:convergence}
        \end{figure}
        \begin{figure}[tb]
            \centering
            \includegraphics[width=0.75\linewidth]{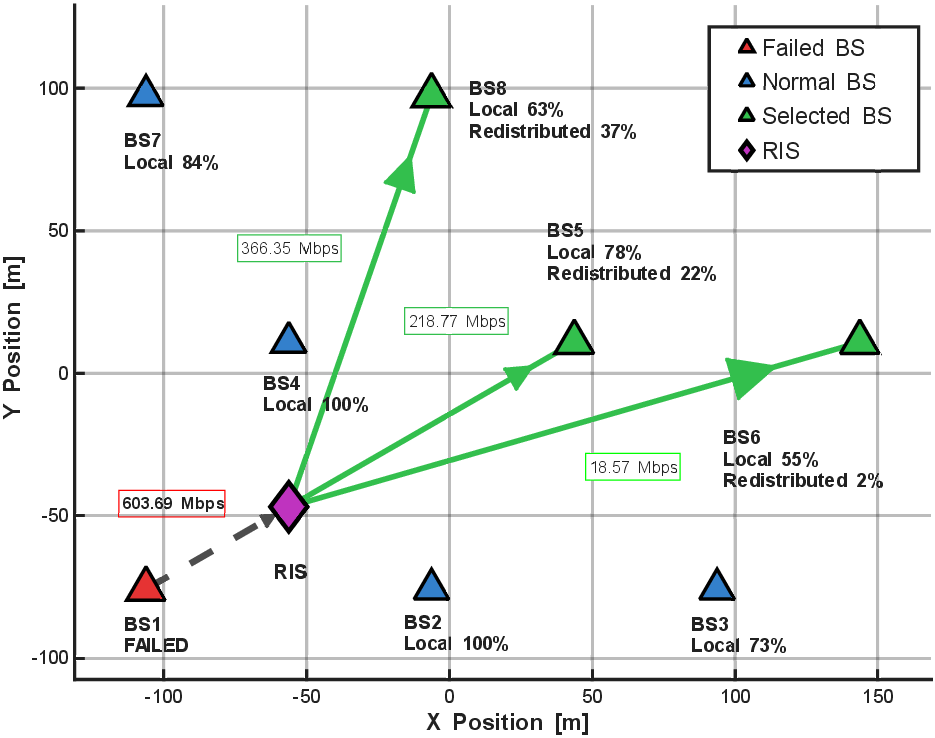}
            \caption{Network snapshot and demonstration of the optimized traffic redistribution. The local traffic $C_l$ in ratio to the total capacity is marked as \emph{Local}. For the selected BSs, the resolvable backhaul traffic $r_l$ in ratio to the total capacity is marked as \emph{Redistributed}.}
            \label{fig:snapshot}
        \end{figure}
        Fig.~\ref{fig:convergence} illustrates the convergence behavior of the proposed algorithm over one network realization. The objective value converges within approximately 20 iterations, demonstrating rapid algorithm convergence. The phase-shift and precoder changes are quantified using $\|\mathbf{\Phi}^{(\varepsilon)}-\mathbf{\Phi}^{(\varepsilon-1)}\|_F$ and $\sum_{l=1}^L\|\mathbf{w}^{(\varepsilon)}_l-\mathbf{w}^{(\varepsilon-1)}_l\|$, respectively. These metrics indicate convergence within 30 iterations, confirming the algorithm's stability. 

        Fig.~\ref{fig:snapshot} visualizes the network deployment and optimized traffic redistribution for one realization. The selected BSs are observed not necessarily to be the closest ones to the disconnected BS, as the optimization accounts for both channel quality and accessible spare capacity. Moreover, as the number of accessible BSs is constrained by the antenna count $N$, suggesting that larger $N$ enables more flexible spare capacity utilization and enhanced survivability. In the following subsections, we investigate survivability under various traffic distributions and antenna configurations.

        \begin{table}[H]
            \caption{System parameters}
            \centering
            \label{tab:coefficient}
            \begin{tabular}{ll|ll}
            \hline
            \textbf{Parameter} & \textbf{Value} & \textbf{Parameter} & \textbf{Value} \\ \hline
            $N$ & 4 & $C_0$ & 1\,Gbps \\
            $M$ & 512 & $\kappa$ & 9\,dB \\
            $L$ & 7 & $B$ & 1\,GHz \\
            $P_{\max}$ & 5\,W & $f_c$ & 28\,GHz \\
            $d_0$ & 100\,m & $\sigma^2$ & $10^{-12}$\,W \\ \hline
            \end{tabular}
        \end{table}

        \subsection{Traffic pattern and survivability}
            
            In practice, traffic conditions across neighboring cells might exhibit strong spatial correlation due to the clustering of user activities and geographical patterns. During peak hours, this spatial correlation intensifies, creating concentrated traffic hotspots where multiple adjacent cells simultaneously experience high utilization. Backhaul link failures occurring within or near these hotspots pose particularly severe challenges, as the affected traffic cannot be easily redistributed to already heavily loaded neighboring BSs, potentially leading to service degradation across the entire hotspot region. 
            
            We model the most challenging scenario where the disconnected BS is near a traffic hotspot, causing nearby BSs with better channel conditions to be heavily loaded. Let $\eta = C_d/C_0$ denote the traffic intensity. The local traffic of surviving BS $l$ is modeled as
            \begin{equation}
                \eta_l = \alpha \left(\frac{d_{\text{max}}-d_{l}}{d_{\text{max}} - d_{\text{min}}}\right)^\gamma\eta + (1-\alpha)\eta+\chi_l,
            \end{equation}
            where $d_{\text{min}}$ and $d_{\text{max}}$ are the minimum and maximum distances from surviving BSs to the disconnected BS, $\gamma\in\mathbb{R}_+$ controls exponential decay, $\chi_l\sim\mathcal{N}(0,\sigma_\chi^2)$ adds stochasticity, and $\alpha\in[0, 1]$ determines the bias strength, with $\alpha\neq0$ indicates the \emph{hotspot} traffic and $\alpha=0$ for \emph{uniform} traffic. 
            
            \begin{figure} [tb]
            \centering
            \includegraphics[width=0.75\linewidth]{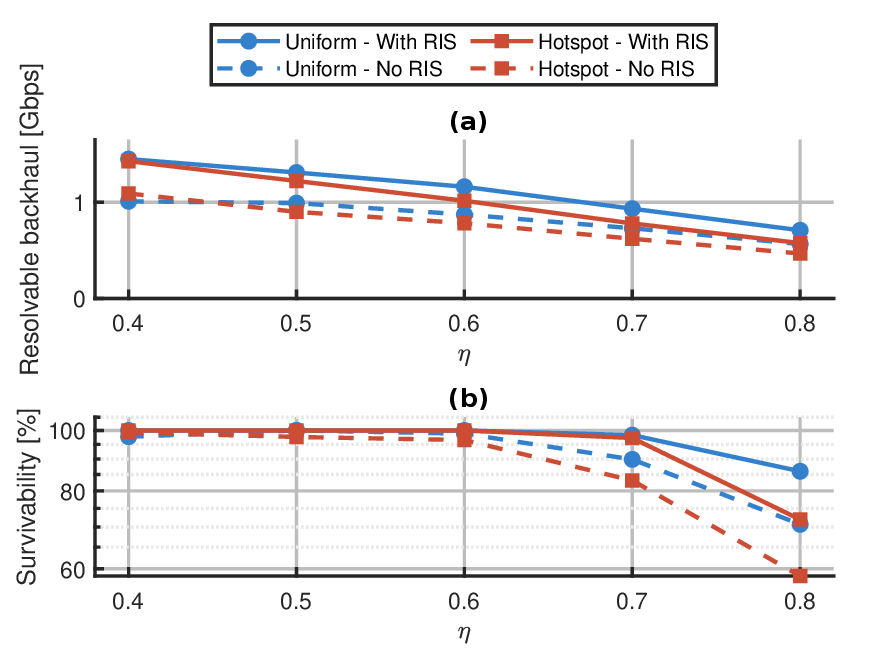}
            \caption{Traffic pattern and corresponding performance. For hotspot cases, $\alpha=0.7$, $\gamma=2$ and $\sigma_\chi = 0.05$. (a) Total resolvable backhaul traffic under various $\eta$; (b) survivability under various $\eta$. For each tested case, 20 instances are optimized, and the average is plotted.}
            \label{fig:loadvarying}
            \end{figure}

            Fig.~\ref{fig:loadvarying}(a) presents the total resolvable backhaul traffic $R$ as a function of the disconnected BS traffic intensity ratio $\eta$. Without RIS assistance, the hotspot traffic scenario yields lower $R$ values compared to the uniform traffic case under a high traffic intensity, highlighting the inherent challenge of resolving traffic when neighboring BSs are already heavily utilized. Based on that, under both traffic scenarios, RIS deployment shows a substantial enhancement in $R$. 
            
            Fig.~\ref{fig:loadvarying}(b) demonstrates the survivability performance across different traffic patterns. At low traffic intensities, the system achieves high survivability even without RIS assistance, as sufficient spare capacity is available across neighboring BSs. The benefits of RIS become most pronounced at spatially correlated traffic under a high intensity ($\eta = 0.8$). RIS deployment improves survivability from $58\%$ to $72\%$, demonstrating its critical role in maintaining network survivability.
        
        \subsection{Number of antennas and survivability}
        
            It is also worth noticing from Fig.~\ref{fig:loadvarying} that the enhancement in $R$ brought by the RIS diminishes as the system-wide traffic intensity increases. This suggests the total accessible spare capacity becomes the limiting factor rather than the wireless channel capacity. As the number of accessible BS is limited by the number of antennas at the disconnected BS, varying the number of antennas will then bring different accessibility to the spare capacity. 
        
            \begin{figure}[tb]
                \centering
                \includegraphics[width=0.75\linewidth]{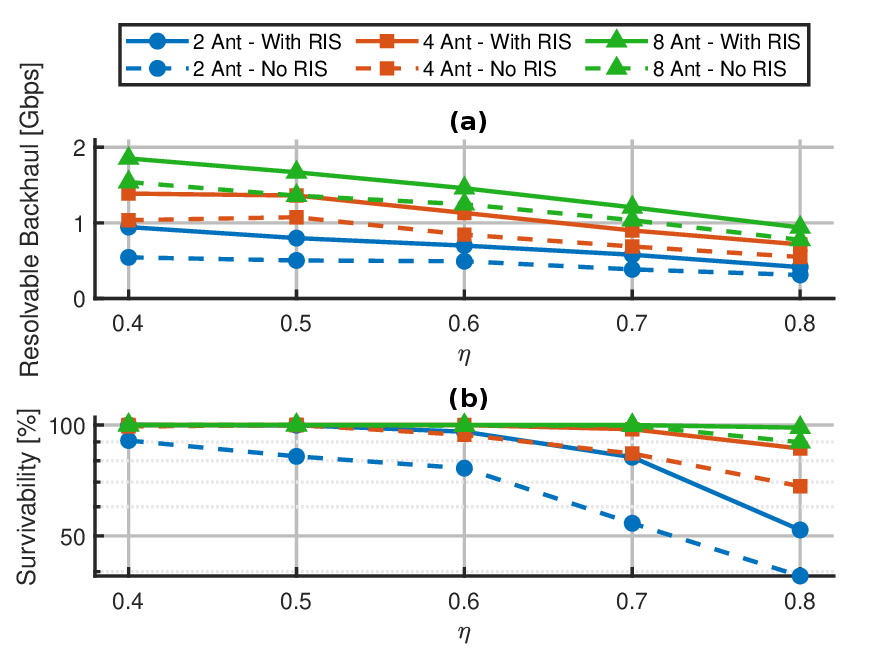}
                \caption{Number of antennas at the BS and corresponding performance under the uniform traffic intensity case. (a) Total resolvable backhaul traffic under various $\eta$; (b) survivability under various $\eta$. For each tested case, 20 instances are optimized, and the average is presented.}
                \label{fig:antennavarying}
            \end{figure}

            As shown in Fig.~\ref{fig:antennavarying}(a), the more antennas at the BSs, the higher $R$ the disconnected BS can achieve. This is also valid when the traffic intensity is low, where the total accessible spare capacity is not limiting. Not only do extra antennas allow access to more BSs, but they also enhance the precoding gain, bringing higher capacity for the wireless traffic redistribution. 

            The RIS provides its most substantial survivability gains when the antenna count is limited, as illustrated in Fig.~\ref{fig:antennavarying}(b). With only two antennas per BS, the survivability suffers from both the spatial diversity limitation and the number of BSs it can access. The RIS is observed to have addressed both constraints simultaneously by extending the signal's effective coverage and creating a rich scattering environment. Consequently, even with a limited number of antennas, the RIS-assisted system achieves nearly $100\%$ survivability under moderate traffic conditions ($\eta \leq 0.6$). 
            
        
    \section{Conclusion}

        The increasing density of small-cell deployments creates both opportunities and vulnerabilities for network survivability. While massive cable infrastructure expansion introduces greater failure risks, high wireless capacity between neighboring BSs enables promising wireless backup solutions. This work addresses single backhaul link failures through RIS-enhanced wireless traffic redistribution,  jointly optimizing target BS selection, precoding vectors, and RIS phase-shift configuration to maximize redistributed traffic and network survivability with the proposed algorithm.

        Our numerical results show that survivability depends critically on both wireless channel capacity and spare backhaul capacity accessibility at other BSs. Under spatially correlated traffic scenarios, RIS deployment improves survivability from $58\%$ to $72\%$ in our case study with high traffic intensity. We reveal that at high traffic loads, total accessible spare capacity, rather than channel capacity, becomes the limiting factor. Additional antennas at BSs have shown an improvement in both spare capacity accessibility and in spatial diversity. For systems with limited antenna counts, RIS shows a substantial gain in enhancing both signal reachability and quality, guaranteeing near $100\%$ survivability under a moderate traffic intensity.

\bibliographystyle{IEEEtran}

\bibliography{Main}

@techreport{epri2023underground,
  title={Failure {R}ates of {U}nderground {C}able {S}ystems},
  author={{Electric Power Research Institute}},
  institution={EPRI},
  year={2023},
  url={https://distribution.epri.com/underground/public/failures/},
  type={Technical Report}
}

@techreport{gallagher2023undersea,
  title={Protection of {U}ndersea {T}elecommunication {C}ables: Issues for Congress},
  author={Gallagher, Jill C. and Carter, Nicole T.},
  institution={Congressional Research Service},
  number={R47648},
  year={2023},
  month={August},
  day={7},
  url={https://www.congress.gov/crs-product/R47648},
  type={CRS Report}
}

@article{donegan2013mobile,
  title={Mobile network outages \& service degradations: A heavy reading survey analysis},
  author={Donegan, Patrick},
  journal={Firmenschrift. Heavy Reading},
  year={2013}
}

@ARTICLE{10685402,
  author={Kagai, Francis and Branch, Philip and But, Jason and Allen, Rebecca and Rice, Mark},
  journal={IEEE Access}, 
  title={{R}apidly {D}eployable {S}atellite-{B}ased {E}mergency {C}ommunications {I}nfrastructure}, 
  year={2024},
  volume={12},
  number={},
  pages={139368-139410},
  doi={10.1109/ACCESS.2024.3465512}}

@ARTICLE{9599638,
  author={Wang, Yuntao and Su, Zhou and Zhang, Ning and Fang, Dongfeng},
  journal={IEEE Wireless Communications}, 
  title={Disaster {R}elief {W}ireless {N}etworks: {C}hallenges and {S}olutions}, 
  year={2021},
  volume={28},
  number={5},
  pages={148-155},
  doi={10.1109/MWC.101.2000518}}

@ARTICLE{6979954,
  author={Gomez, Karina and Goratti, Leonardo and Rasheed, Tinku and Reynaud, Laurent},
  journal={IEEE Communications Magazine}, 
  title={Enabling disaster-resilient {4G} mobile communication networks}, 
  year={2014},
  volume={52},
  number={12},
  pages={66-73},
  doi={10.1109/MCOM.2014.6979954}}

@INPROCEEDINGS{11046130,
  author={Manzoor, Aunas and Ozger, Mustafa and Cavdar, Cicek},
  booktitle={2025 21st International Conference on the Design of Reliable Communication Networks (DRCN)}, 
  title={Risk-{A}ware {B}ackup {P}ath {A}llocation in {O-RAN} {B}ased {I}ntegrated {T}errestrial and {N}on-{T}errestrial {N}etworks}, 
  year={2025},
  volume={},
  number={},
  pages={1-6},
  doi={10.1109/DRCN65040.2025.11046130}}

@article{barlacchi2015multi,
  title={A multi-source dataset of urban life in the city of {M}ilan and the {P}rovince of {T}rentino},
  author={Barlacchi, Gianni and De Nadai, Marco and Larcher, Roberto and Casella, Antonio and Chitic, Cristiana and Torrisi, Giovanni and Antonelli, Fabrizio and Vespignani, Alessandro and Pentland, Alex and Lepri, Bruno},
  journal={Scientific data},
  volume={2},
  number={1},
  pages={1--15},
  year={2015},
  publisher={Nature Publishing Group}
}

@ARTICLE{8466626,
  author={Wang, Xu and Zhou, Zimu and Xiao, Fu and Xing, Kai and Yang, Zheng and Liu, Yunhao and Peng, Chunyi},
  journal={IEEE Transactions on Mobile Computing}, 
  title={Spatio-{T}emporal {A}nalysis and {P}rediction of {C}ellular {T}raffic in {M}etropolis}, 
  year={2019},
  volume={18},
  number={9},
  pages={2190-2202},
  doi={10.1109/TMC.2018.2870135}}

@article{shen2018fractional,
  title={Fractional programming for communication systems—{Part I}: Power control and beamforming},
  author={Shen, Kaiming and Yu, Wei},
  journal={IEEE Transactions on Signal Processing},
  volume={66},
  number={10},
  pages={2616--2630},
  year={2018},
  publisher={IEEE}
}

@misc{mosek,
  author = {{MOSEK ApS}},
  title = {{The MOSEK Optimization Toolbox for MATLAB}},
  howpublished = {\url{https://www.mosek.com/}},
  year = {2024},
}

@inproceedings{sun2016propagation,
  title={Propagation path loss models for {5G} urban micro-and macro-cellular scenarios},
  author={Sun, Shu and Rappaport, Theodore S and Rangan, Sundeep and Thomas, Timothy A and Ghosh, Amitava and Kovacs, Istvan Z and Rodriguez, Ignacio and Koymen, Ozge and Partyka, Andrzej and Jarvelainen, Jan},
  booktitle={2016 IEEE 83rd Vehicular Technology Conference (VTC Spring)},
  pages={1--6},
  year={2016},
  organization={IEEE}
}

@article{mu2021simultaneously,
  title={Simultaneously transmitting and reflecting ({STAR}) {RIS} aided wireless communications},
  author={Mu, Xidong and Liu, Yuanwei and Guo, Li and Lin, Jiaru and Schober, Robert},
  journal={IEEE transactions on wireless communications},
  volume={21},
  number={5},
  pages={3083--3098},
  year={2021},
  publisher={IEEE}
}

@ARTICLE{10187713,
  author={Reifert, Robert-Jeron and Roth, Stefan and Ahmad, Alaa Alameer and Sezgin, Aydin},
  journal={IEEE Transactions on Vehicular Technology}, 
  title={{Comeback Kid: Resilience for Mixed-Critical Wireless Network Resource Management}}, 
  year={2023},
  volume={72},
  number={12},
  pages={16177-16194},
  doi={10.1109/TVT.2023.3296977}}

@article{shi2024ris,
  title={{RIS}-aided cell-free massive MIMO systems for 6G: {F}undamentals, system design, and applications},
  author={Shi, Enyu and Zhang, Jiayi and Du, Hongyang and Ai, Bo and Yuen, Chau and Niyato, Dusit and Letaief, Khaled B and Shen, Xuemin},
  journal={Proceedings of the IEEE},
  volume={112},
  number={4},
  pages={331--364},
  year={2024},
  publisher={IEEE}
}

@article{townend2023challenges,
  title={Challenges and opportunities in wireless fronthaul},
  author={Townend, Dave and Husbands, Ryan and Walker, Stuart D and Sutton, Andy},
  journal={IEEE Access},
  volume={11},
  pages={106607--106619},
  year={2023},
  publisher={IEEE}
}

@book{zhang2021reconfigurable,
  title={Reconfigurable intelligent surface-empowered {6G}},
  author={Zhang, Hongliang and Di, Boya and Song, Lingyang and Han, Zhu},
  year={2021},
  publisher={Springer}
}

@inproceedings{huang2021towards,
  title={Towards reliable communications in intelligent reflecting surface-aided cell-free {MIMO} systems},
  author={Huang, Rui and Wong, Vincent WS},
  booktitle={2021 IEEE Global Communications Conference (GLOBECOM)},
  pages={1--6},
  year={2021},
  organization={IEEE}
}

\end{document}